\documentclass{chi2012}
\pdfoutput=1

\usepackage{balance}  
\usepackage{graphics} 
\usepackage{times}    
\usepackage{url}      

\usepackage[utf8]{inputenc}
\usepackage[english]{babel}

\usepackage{amssymb, amsmath}
\usepackage{IEEEtrantools}  
\usepackage{graphicx}
\usepackage{subfig}   
\usepackage[table]{xcolor}  
\usepackage{listings}  

\usepackage[colorinlistoftodos]{todonotes}

\makeatletter
\def\url@leostyle{%
  \@ifundefined{selectfont}{\def\UrlFont{\sf}}{\def\UrlFont{\small\bf\ttfamily}}}
\makeatother
\def\pprw{8.5in}
\def\pprh{11in}

\usepackage[pdftex]{hyperref}
\hypersetup{
pdftitle={Assessing the Performance of Question-and-Answer Communities Using Survival Analysis},
pdfauthor={LaTeX},
pdfkeywords={Question-and-answer websites; Stack Exchange; performance metrics; survival
    analysis},
bookmarksnumbered,
pdfstartview={FitH},
colorlinks,
citecolor=black,
filecolor=black,
linkcolor=black,
urlcolor=black,
breaklinks=true,
}

\newcommand\tabhead[1]{\small\textbf{#1}}

\begin{document}

\title{Assessing the Performance of Question-and-Answer Communities Using Survival Analysis}

\numberofauthors{4}
\author{
  \alignauthor Felipe Ortega\\
    \affaddr{DIS, University Rey Juan Carlos}\\
    \affaddr{Mostoles, Spain.}\\
    \email{felipe.ortega@urjc.es}
  \alignauthor Gregorio Convertino\\
    \affaddr{Informatica Corporation}\\
    \affaddr{Redwood City, CA, USA}\\
    \email{gconvertino@informatica.com}
  \alignauthor Massimo Zancanaro\\
    \affaddr{IRST, Fondazione Bruno Kessler}\\
    \affaddr{Trento, Italy}\\
    \email{zancana@fbk.eu}
  \and
  \alignauthor Tiziano Piccardi\\
    \affaddr{University of Trento}\\
    \affaddr{Trento, Italy}\\
    \email{piccardi@disi.unitn.it}
}

\maketitle

\begin{abstract}
Question-\&-Answer (QA) websites have emerged as efficient platforms for knowledge sharing and problem solving. In particular, the \textit{Stack Exchange} platform includes some of the most popular QA communities to date, such as \textit{Stack Overflow}. Initial metrics used to assess the performance of these communities include summative statistics like the percentage of resolved questions or the average time to receive and validate correct answers. However, more advanced methods for longitudinal data analysis can provide further insights on the QA process, by enabling identification of key predictive factors and systematic comparison of performance across different QA communities. In this paper, we apply survival analysis to a selection of communities from the \textit{Stack Exchange platform}. We illustrate the advantages of using the proposed methodology to characterize and evaluate the performance of QA communities, and then point to some implications for the design and management of QA platforms.  
\end{abstract}


\category{H.3.4.}{Systems and Software}{Performance evaluation (efficiency and effectiveness)}


\terms{
	Design; Measurement; Performance.
}


\section{Introduction}

Collaborative knowledge generation and problem solving have been 
transformed by social technologies over the past decade. The arrival
of Web 2.0 technologies unleashed the true potential of knowledge sharing
platforms that are powered by large-scale human participation. 
Traditional platforms for knowledge sharing and peer support
such as mailing lists~\cite{kuk2006} have been eventually superseded by more 
interactive and dynamic environments such as forums, blogs and 
wikis~\cite{wagner2004}. Thus, a new area of big data analytics and 
application has emerged with the aim to understand and support such large-scale processes: 
see the growing number of studies on Wikipedia, Twitter, and social networking software tools.

More recently, Question-\&-Answer (QA) websites such as those comprising 
the \textit{Stack Exchange Network} introduced new ways to improve the 
efficiency of problem-based sharing and collaboration in distributed communities.
For instance, the inclusion of interesting \textit{gamification} incentives~\cite{deterding2011} contributes to increase user engagement: respondents are 
granted a reputation score and a badge that reflect their speed to solve questions 
and the number of votes received by their answers.

Some of these communities reached enough popularity to become \textit{the} 
dominant platforms for worldwide knowledge exchange and problem solving in
specialized domains~\cite{vasilescu2014}. Such is the case of \textit{StackOverflow}, 
aimed at programmers, or \textit{Cross Validated}, for people who must solve problems in
statistics, machine learning, and data analytics. In fact, the former 
has been acclaimed as the "fastest QA site in the West" in a previous 
research study~\cite{mamykina2011}, featuring more than 92\% of its
total number of questions answered in a median time of just 11 minutes. 

Given these promising results, the utilization of QA platforms
as a replacement of traditional help desk services in specialized
domains increasingly appears a viable alternative. To test the feasibility of this alternative, it is essential that we can measure the performance of QA platforms using suitable data analytics methods. \textit{Stack Exchange} regularly publishes 
detailed datasets tracking the question answering process 
in their 115 communities. Therefore this platform represents an excellent testbed to study 
the efficiency of QA platforms.

Studies evaluating the performance of these communities have
focused predominantly on \textit{descriptive} metrics (e.g. mean/median answering time, proportion 
of solved questions, etc.), as in~\cite{mamykina2011}. However, more advanced methods for \textit{longitudinal} data 
analysis are required to integrate temporal information about the question 
resolution process as an integral part of the evaluation model. \textit{Survival analysis} provides methodology and statistical 
techniques for the analysis of time-to-event data~\cite{hosmer2011} that can
fill the current methodological gap. In our application of this analysis, we define the event of interest as the "time elapsed until a question is resolved". 

Survival analysis has already been applied in big data analytics for web 
systems, such as modeling conversions in online advertising~\cite{chandler2010}.
The method can also incorporate valuable information about \textit{unresolved} 
questions, which have been often been overlooked in prior studies. But, predicting unanswered questions is a critical goal of our work because we are testing the applicability of QA platforms for offering reliable 
customer support service in specialized domains: i.e., while having 8\% 
of unanswered questions is acceptable for the \textit{StackOverflow} site, the same would not be admissible for a company that sells this service to business customers. 

Hence, the aim of this paper is to introduce the use of survival analysis
to model the occurrence of events of interest in online communities and web systems
through the analysis of big datasets. This will lead us to: 
\begin{itemize}
\item Characterize and compare the performance of different QA sites in which an 
event of interest occurs;
\item Identify relevant factors that exert a positive or negative influence over 
the happening of such event of interest; 
\item Estimate the expected time of occurrence of future events of interest.
\end{itemize}

We analyze the performance as time to answer in eight different sites on the 
\textit{Stack Exchange} platform. Through this study, we first illustrate the 
applicability of survival analysis for big data analytics and then draw implications 
for the design and management of QA platforms.

\section{Related research}
QA sites are a Web 2.0 technology increasingly adopted by self-help communities of experts 
such as programmers, mathematicians, and statisticians. As a result, these sites are 
changing the ways in which these communities share knowledge and collaborate on problem solving. 
For example, Vasilescu and collaborators found in~\cite{vasilescu2014} that the emergence 
of a QA community in Stack Exchange is causing experts in a specific community to migrate 
from an existing mailing list (\textit{r-help}) to a Stack Exchange site (as part of StackOverflow), 
where their behaviour is different. The migration appears motivated by the incentives 
and gamification mechanisms applied in Stack Exchange to engage the users. Tausczik et
al.~\cite{tausczik2014} present qualitative findings on the uses of MathOverflow by
mathematicians as a large-scale problem-solving platform. While there is evidence of 
new emerging practices, however, most of the studies have focused on individual QA 
communities, which limits generalizability, and contributed findings that are often 
anecdotal or descriptive rather than predictive in nature.

Despite the current methodological limitations, QA sites (and Stack Exchange in particular) 
are receiving increasing attention from researchers as big data resources for 
studying community phenomena and design future technologies, as witnessed by the 
growing number of papers published each year on QA sites 
(e.g., http://meta.stackoverflow.com/q/134495). 

The studies have focused on a variety of facets of the QA communities. Several of the 
early investigations have focused on modeling expertise of users~\cite{zhang2007,adamic2008}.
But more recently there is growing interest for characterizing the principles that 
regulate the process of QA~\cite{anderson2012}. Some studies started analyzing the 
factors that may predict user intent~\cite{chen2012}; the quality of the answer~\cite{wang2013}
and the likelihood of getting an answer~\cite{treude2011}. In our own previous work, 
we have analyzed the impact of some non content-related characteristics of the 
question to estimate the likelihood that a given question will receive a satisfactory 
answer in a reasonable time~\cite{piccardi2014}. In turn, Tausczik et al. try to 
predict the perceived quality of online mathematics contributions from users' 
reputations~\cite{tausczik2011,tausczik2012}.

Arguably the most complete report on StackOverflow up to 2011, ~\cite{mamykina2011} 
characterized the site’s properties and evolution that contributed to its 
success. Focusing on questions that eventually received answers, they reported a 
median time for first answers of 11 minutes and a median time for accepted answers of 21 
minutes, which are consistent with our subsequent analysis~\cite{piccardi2014}.

Although for some respect different, Yahoo! Answers is another commonly studied QA site. Differently from the Stack Exchange, this site is for general-purpose questions. It had 
60 million unique visitors and 160 millions answers within the first year~\cite{pal2012}.
Between 2011 and 2012, it attracted between 17 and 24 million unique visitors per month.
Yahoo! Answers has attracted a relevant number of studies too. In~\cite{chen2012}, Chen 
and colleagues classify questions in three categories according to their underlying user 
intent: subjective, objective, and social. They build a predictive model through machine
learning based on both text and metadata features (topic, time, user expertise). Wang et
al.~\cite{wang2013} point to the problem of low-value questions introducing noise as 
sites grow: they report signs of stalling in the user growth of Yahoo! Answers, with 
traffic dropping 23\% in a span of four months in 2011.

\section{Methodology and definitions}

In this section, we first describe the public data sources that we have used. Then, we introduce the essential elements of survival analysis required to
understand and interpret the results of the study. Finally, we introduce the features used in order to model the time to answer a question.

\subsection{Data sources: Stack Exchange Data Dumps}

Every 3 months, \textit{Stack Exchange} publishes anonymized dumps of all content stored in
databases of their public QA sites (115 different communities, to date). The data 
is hosted by the Internet Archive project (\url{https://archive.org/details/stackexchange}) 
and it is available for direct download or using the BitTorrent file sharing protocol. 
The dump files include information in XML format (except for users’ account data) for 
each site in the platform, or network, and it can be used for research purposes. 

Data included in these dump files characterize question threads: comments, answers, answers marked as accepted, votes received by the answers, badges 
and special user levels granted to users based on their question-solving merits, etc.
The datasets that we analyzed were released in March 2013 and included the most up-to-date version
of each question thread along with the history of each post (edits, votes, etc). 

We used a relational database (PostgreSQL) to rebuild the original data model and 
for computing new aggregate information. We created some intermediate tables for 
an incremental analysis during the features selection stage. In this phase, to select 
the most relevant features we generated descriptive statistics for each aspect of the
communities using the R programming language. The list of relevant features considered
will be described below.

In particular, we have focused the scope of this analysis on a subset of eight communities in the \textit{Stack Exchange} network, representing a variety of application domains:

\begin{itemize}

	\item \textit{Apple (AskDifferent)}: Apple users and developers.
    \item \textit{AskUbuntu}: Ubuntu users and developers.
    \item \textit{Math (Mathematics)}: Math students, instructors, and 
    professionals  in this field.
    \item \textit{ServerFault}: system and network administrators.
    \item \textit{SharePoint}: SharePoint users and developers.
    \item \textit{StackOverflow}: professional and enthusiast programmers.
    \item \textit{SuperUser}: computer enthusiasts and power users.
    \item \textit{Wordpress (Wordpress Answers)}: Wordpress administrators and developers.

\end{itemize}

To undertake this study, we took a random sample of 5,000 questions from each site.
In each sample, we filtered out all cases of questions showing obvious wrong values for 
the resolution datetime field (for instance, lower than the creation datetime). These 
values were due to spurious inconsistencies in the creation of dump files.

Questions can receive several answers and site users can cast votes on these answers to
help indicate which are, in their opinion, the most useful ones. However, only the original 
user who posted a question can  finally mark it as \textit{resolved}, by picking up one of 
the answers as \textit{accepted}. This may or may not coincide with the answer that received 
the highest number of votes. Therefore, the voting process provides an alternative metric 
to assess the respondents' reputation level. Questions marked as off-topic and removed by 
site moderators have been filtered out of our analysis. 

In our case, we are interested in modeling the \textit{time elapsed until each 
question is resolved} in the eight communities included in the study. To this aim, we 
created a dummy variable \texttt{status} to identify answers that were marked as
\textit{accepted} from those still waiting for resolution (even if they may have
already received answers). Then, we apply survival 
analysis to create a model for the time elapsed until questions are resolved, using 
this binary indicator to identify our event of interest. A key advantage
of survival analysis is that it can also include information about unresolved questions (censored)
in the model, which leads to more accurate estimations. We introduce this and
other details about survival analysis in the following subsection.

\subsection{Survival Analysis}

We introduce here a collection of statistical methods and techniques to handle 
timing and duration until a certain event of interest occurs. Although these techniques
are frequently referred to as \textit{survival analysis} 
in disciplines like medicine, biostatistics 
and epidemiology, they are also well-known in other scientific areas under different names~\cite{mills2011}:
reliability analysis (engineering and production); duration model (economics) and
event history analysis (social sciences). In spite of this, these techniques are rarely applied in computer science. We could find only a few studies that applied survival analysis: e.g., to model the retention of contributors in open collaborative projects
\cite{ortega2009, zhang2012}, gender imbalance in Wikipedia \cite{lam2011} or the
duration of open source projects \cite{samoladas2010}.

The goal of survival analysis is to model the \textit{hazard rate}, that is, the 
conditional probability that an event of interest occurs at a specific time interval 
$t$. Let $T$ be a random variable representing the time until the event of interest 
happens which,in our case, will be "\textit{time until a question is resolved}" 
(measured in minutes). 
Thus, the \textit{hazard rate} represents the rate at which questions in our study
experiment this event at time $t$, conditional on surviving (not experiencing the event) 
up to that time:

\begin{equation}
\label{eq:hazard-function}
  h(t) = \lim_{\Delta t \to 0} \frac{Pr\left[\left(t \leq T < t + \Delta t\right)|T\geq t\right]}{\Delta t}
\end{equation}

Hence, in equation~(\ref{eq:hazard-function}), $h(t)$ represents the instantaneous risk 
that the event of interest happens in the interval $[t, t+\Delta t]$. Along with the
hazard rate, it is frequent to consider the \textit{survival function}, $S(t)$, which 
indicates the probability that the survival time $T$ is greater or equal than a given 
time $t$. Both functions are related in the following way:

\begin{equation}
  h(t)=\frac{f(t)}{S(t)}
\end{equation}

Where $f(t)$ is the unconditional probability density function of the random variable $T$,
representing survival time. For descriptive purposes, a non-parametric method known as the
\textit{Kaplan-Meier} (KM) estimator~\cite{kaplan-meier1958} is usually applied to obtain 
a graphical representation of the survival function.

A remarkable feature of survival analysis is that it can accommodate \textit{censored data}. 
In the most frequent cases, either the observation period expires or a subject is removed 
from the study before the event of interest occurs. In these cases, we have some information
about survival time, but not the exact value of that survival time. All we know is that 
the event of interest did not occur (yet) by the end of the study. This is called 
\textit{right-censoring} and it is the only form of censoring accounted for in this analysis.
Thus, right-censored cases correspond to questions that either not received any
answer or, despite receiving any answer, it has not been accepted by the author of the question.

This offers a key advantage for the study of time-to-event data (Hosmer, 2007), as we 
are not forced to make any assumptions about the status of questions that could be 
eventually resolved in the future, after the end of the observation period. As a result, predictions  about the expected resolution time 
for new questions are likely to be less biased.

One of the most popular models in survival analysis is the \textit{Cox proportional-hazards
model} (Cox PH). In this model, the hazard rate is represented by:

\begin{equation}
    h_{i}(t) = h_{0}(t)\: exp\left(\beta_{1}x_{i1}+\beta_{2}x_{i2}+\cdots+\beta_{k}x_{ik}\right)
\end{equation}

Where $h_{0}(t)$ represents a baseline hazard function that remains
unspecified, $x_{ik}$ are $k$ fixed covariates for each observation
$i$, and $\beta_{k}$ are $k$ regression coefficients. The model is usually expressed
in terms of the \textit{hazard ratio} of two given observations, $i$ and $j$:

\begin{equation}
    log\left\{\frac{h_{i}(t)}{h_{j}(t)}\right\} = \beta_{1}x_{i1}+\beta_{2}x_{i2}+\cdots+\beta_{k}x_{ik}
\label{eq:hazard-ratio}
\end{equation}

Similarly to the log-odds ratio in logistic regression models, the
hazard ratio is easy to interpret in this case, since the baseline
hazard function $h_{0}(t)$ cancels out in the numerator and denominator of
equation~\ref{eq:hazard-ratio}. Therefore, the hazard ratio between any
two observations is independent of time. This is why the
Cox model is called a \textit{proportional hazards} model. We use it
to identify any covariates that influence (in a positive or
negative way) the resolution time of questions. Once the model is specified, 
we can obtain an estimation of the influence of each covariate on the hazard
function. Likewise, we can also predict the expected survival time for new 
cases.

In spite of these advantages, this standard formulation of the Cox PH model 
assumes that all covariates enter the model in a linear fashion. However, it 
is quite frequent that the log hazard ration (output variable) does not depend 
linearly on some of the model's covariates. In this case, it is necessary 
to relax the linearity assumptions for model parameters, as explained 
in section 2.4 of~\cite{harrell2001}. A practical alternative to achieve this
goal is using the so-called restricted cubic splines functions with $k$ knots,
introduced by Stone and Koo in~\cite{stone1985}, to transform any covariates we 
may suspect that do not follow a linear relationship with the outcome variable.

The library \texttt{rms} for the R statistical programming language~\cite{Rlang2014} 
implements many advanced techniques explained in~\cite{harrell2001} and provides
support for this transformation on independent parameters with the \texttt{rcs()}
function (along with other possible alternatives). For an independent parameter
$X$ in the model, the restricted cubic spline function with $k$ knots 
$t_{1},\dots,t_{k}$ is given
by~\cite{devlin1986}:

\begin{IEEEeqnarray}{rCl}
f(X) & = & \displaystyle\beta_{0} \,+\, \beta{1}X \,+\, \beta{2}(X-t_{1})^{3} \,+\, \beta_{3}(X-t_{2})^{3} \,+\nonumber\\
&& +\> \dots \,+\,\beta_{k+1}(X-t_{k})^{3}
\end{IEEEeqnarray}

Moreover, additional functions included in the \texttt{rms} package let users represent 
the effects of covariates graphically taking these transformations into account,
so that results are plotted back on the original scale of the model's covariates to
facilitate their intuitive interpretation. In the same way, multilevel confidence 
intervals can be computed to visually assess the size of effects of independent 
parameters on the log hazard ratio, according to modern strategies for informative 
and robust evaluation of statistical models~\cite{cumming2013}. We will make use 
of these features to present the results of our survival model fitting on the 
Stack Exchange sites under study in the following sections.

\subsection{Relevant features to model question answering}
\label{subsec:model-features}

Table~\ref{tab:features-list} summarizes the list of features that we include as
covariates to model the resolution time of questions in the eight communities analyzed 
with the the Cox PH model.

\begin{table}[h!]
  \centering
  \begin{tabular}{|c|c|l|}
    \hline
    \tabhead{Feature} &
    \multicolumn{1}{|p{0.2\columnwidth}|}{\centering\tabhead{Type}} &
    \multicolumn{1}{|p{0.4\columnwidth}|}{\centering\tabhead{Description}} \\
    \hline
    bodylength & Integer & \multicolumn{1}{|p{0.4\columnwidth}|}{Number of
    printable characters in the body of questions (HTML filtered out).}\\
    \hline
    titlelength & Integer & \multicolumn{1}{|p{0.4\columnwidth}|}{Number of
    printable characters in the title of questions (HTML filtered out).}\\
    \hline
    hasexample & Boolean & \multicolumn{1}{|p{0.4\columnwidth}|}{Dummy variable, 
    indicates if the question contains an example (e.g. code).}\\
    \hline
    tagscount & Integer & \multicolumn{1}{|p{0.4\columnwidth}|}{Number of tags used 
    in the question, for content classification (values in $[2,5]$).}\\
    \hline
    sumpeople & Integer & \multicolumn{1}{|p{0.4\columnwidth}|}{The sum of the
    sizes of the tag-based communities of respondents, considering all the tags of the
    question. We compute in advance the number of active contributors for each tag.}\\
    \hline
    zscore & Decimal & \multicolumn{1}{|p{0.4\columnwidth}|}{Normalized ratio 
    of the difference between questions and answers posted. That is, answering a greater
    proportion of questions (relative to one’s own activity) implies higher expertise.}\\
    \hline
  \end{tabular}
  \caption{List of relevant features to be included in the survival model for
  time to answer questions.}
  \label{tab:features-list}
\end{table}

These six features have been included in our model based on theoretical
background supporting their relationship with the question answering process.

The \textit{body\_length} is the number of printable characters in the body of the 
question after filtering out the HTML formatting. It is considered important 
since unanswered questions are sometimes too short~\cite{treude2011} or, on 
the contrary, they may be too long and tedious~\cite{mamykina2011}. The 
\textit{title\_length} is the number of printable characters in the title. Treude 
and colleagues~\cite{treude2011} discussed the importance of the title 
in at least one QA site in attracting the attention of the community. We 
consider the length of a title as a proxy of its content.

\textit{Has\_example} is a Boolean variable that indicates whether the question 
includes an example (for instance, a snippet of programming code). The importance 
of examples have been discussed by several previous studies~\cite{asaduzzaman2013,
mamykina2011,treude2011}. 

\textit{Tagscount} is the number of tags used in the question.
Although this aspect has not been directly discussed in literature, there is 
some evidence supporting that if questioners facilitate a quick understanding of the 
main topic(s) in the content of questions this may increase the likelihood of 
getting answers~\cite{treude2011,asaduzzaman2013}. StackOverflow’s 
guidelines encourage the use of tags (up to 5) exactly to meet this purpose.

\textit{Sum\_people} is a measure of how many persons will be exposed to 
the questions. As discussed by Harper and colleagues~\cite{harper2008} a 
wider audience increases the likelihood for a question to be answered. In the same
way, open collaborative projects are known to benefit from a large a varied
audience (in terms of previous background and expertise) to solve specific
problems, following \textit{the wisdom of crowds} effect explained by
Surowiecki~\cite{surowiecki2005}.

Finally, the \textit{zscore} metric has been introduced by Zhang and 
colleagues in their analysis of the Java Forum~\cite{zhang2007}. It is 
computed as the normalized ratio of the difference between questions and answers 
and it models the understanding that answering a greater proportion of 
questions (relative to one’s own activity) implies higher expertise. 
Hence, this feature is calculated as follows:

\begin{equation}
zscore = \frac{|a|-|q|}{\sqrt{|a|+|q|}},
\label{eq:zscore}
\end{equation}

where $a$ is the number of answers and $q$ is the number of questions posted by the user.
Since this value changes in time, for each question we used the posts history to compute 
the zscore of the author at the posting time. For each question and answer, we assigned 
an incremental counter to keep track of their temporal position.\\

\section{Results}

In this section, we lay out the main results of our study. In the first place, we evaluate
the performance of the different sites regarding the time to solve questions comparing the
estimators of their corresponding survival functions. This is followed by the results 
of fitting a Cox PH model for each site, using the 6 relevant covariates presented 
above whose role is supported by previous research in this area.

\subsection{Comparing the performance of QA communities}

\begin{figure}[!h]
\centering
\includegraphics[width=0.95\columnwidth]{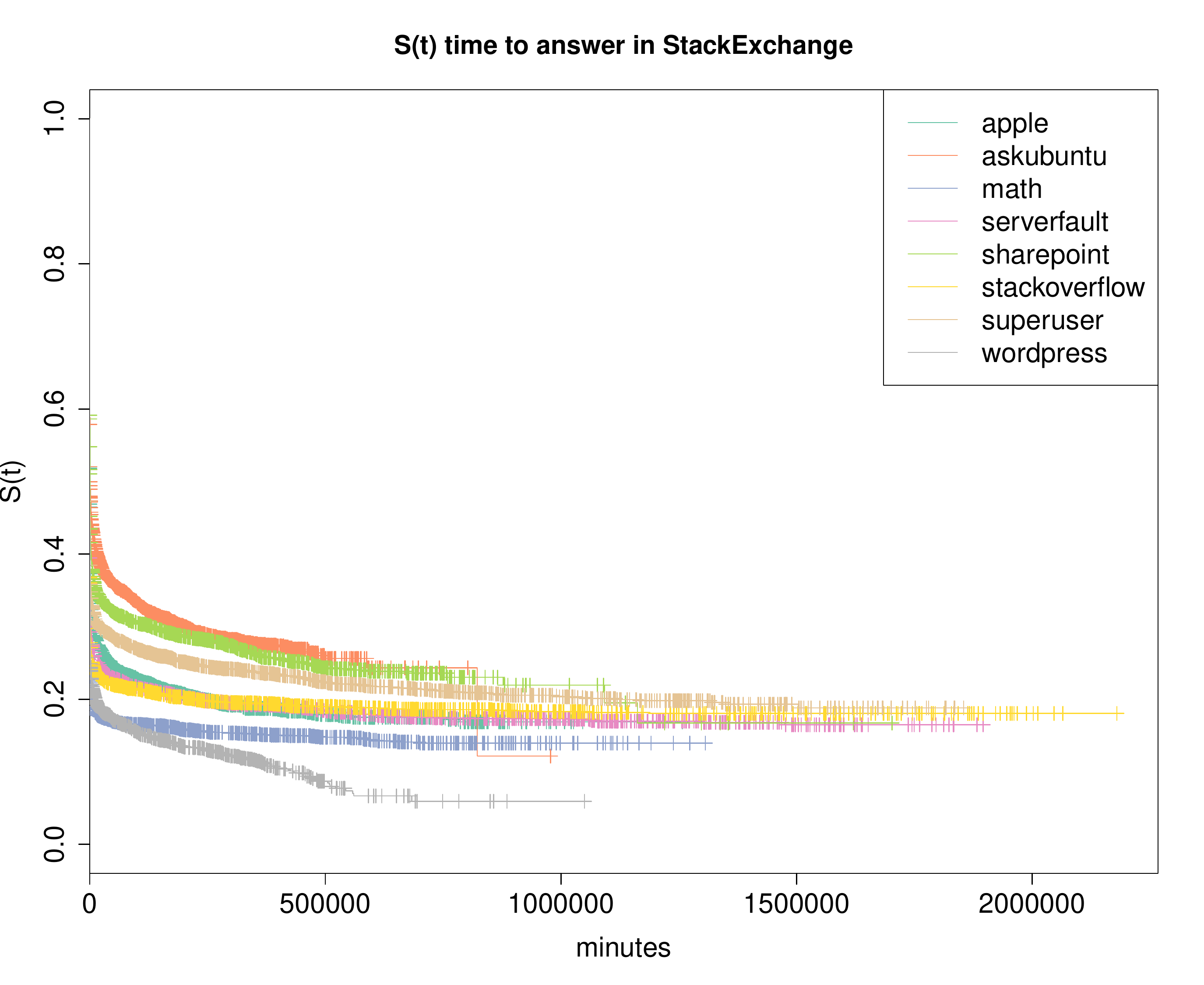}
\caption{Non-parametric Kaplan-Meier estimators of survival functions for time to answer questions in 8 communities from the \textit{Stack Exchange} network}
\label{fig:km-all-sites}
\end{figure}

Now, we present the results of a descriptive analysis of the time to answer questions, 
using the Kaplan-Meier non-parametric estimator. Figure \ref{fig:km-all-sites} 
depicts the estimated survival function $S(t)$ for each site. As we
introduced above, this function is directly linked to the hazard rate. It indicates the
absolute probability that a survival time $T$ is greater or equal than some time $t$
\cite{mills2011}. Therefore, this function represents the proportion of subjects surviving
beyond time $t$, and at $t=0$, $S(0)=1$. In our study, the curve for each site represents
the proportion of questions that remained unsolved for $T \geq t$.

Visual inspection suggests common patterns but also performance differences among the eight communities under study. In all cases, the proportion of questions that remain unsolved
quickly drops below 40\%, showing a good performance on question resolution for all
sites. However, the main differences among sites emerge from their efficiency dealing with
questions that remain unsolved for longer periods of time. For instance, these 
correspond to questions that may be harder to solve (for some reason) or that do not 
attract enough attention from the respondents audience. 

Listing 1 presents the descriptive summary for each site, created
with the \texttt{survival} package in the R programming language ~\cite{survival-package}.
We can see the total number of questions in each sample (after removing 
obvious erroneous cases), the median survival time and 95\% 
confidence intervals for this median value.

\begin{lstlisting}[breaklines=true,
caption={Summary of number of events (solved questions), median 
survival time (in minutes) and its associated 95\% confidence interval 
for each site.},
basicstyle=\ttfamily\scriptsize,
frame=lines,
showtabs=false,
showspaces=false,
showstringspaces=false,
captionpos=b]
> stackex_surv = with(survdata, Surv(tanswer, solved))
> stackex_fit = with(survdata, survfit(stackex_surv ~ site))
> print(stackex_fit)
Call: survfit.formula(formula = stackex_surv ~ site)

              records events median 0.95LCL 0.95UCL
apple            4296   3416  210.0   186.0   247.3
askubuntu        3983   2759  904.4   607.1  1208.9
math             4744   3999   39.6    35.9    43.2
serverfault      4552   3702   84.6    73.8    96.2
sharepoint       4335   3137  417.7   343.3   505.0
stackoverflow    4779   3844   35.8    31.9    42.6
superuser        4467   3433  100.8    87.9   116.3
wordpress        4187   3617  109.5    97.9   120.4
\end{lstlisting}

In particular, the median survival time at which 50\% of all questions in 
each sample are resolved is
estimated by tracing an horizontal line at 0.5 on the plot of the survival curve. 
Both solved questions (with an accepted answer) and unresolved ones (right-censored 
cases) are included in the non-parametric estimation of the survival curve. As we
can see, the performance on question resolution is quite different in each site.
While \textit{StackOverflow} or \textit{Math} present low median survival
times of slightly more than 30 minutes, \textit{AskUbuntu} needs more than
15 hours (904.4 minutes) to resolve half of the total number of questions in the
sample.

A Mantel-Haenszel test \cite{mantel1959} to formally check for statistically
significant differences of survival curves among sites returned a Chi-square value
of 1084 on 7 degrees of freedom, with a virtually null p-value. Hence, we can conclude
that there exist significant differences in the ability of different sites in the
study to resolve questions. 

\subsection{Modeling question answering processes}

As we introduced above, the Cox proportional hazards model (Cox PH) let us create
survival models for inference and prediction of hazard rates, without forcing us
to choose a predefined parametric form of the baseline hazard function. In this
study, we use the six features presented above as parameters of our Cox PH model. 

We first transform the continuous parameters
\texttt{bodylength}, \texttt{titlelength} and \texttt{sumpeople} by taking the natural 
logarithm (\texttt{log}) to reduced initial skewness in their distribution of values.
Then, we apply a restricted cubic spline with 3 knots (\texttt{rcs(x, 3)}) to all
parameters except for \texttt{hasexample} (a binary categorical parameter) and
\texttt{tagscount} (restricted to the interval $[2, 5]$), as we suspect that all
other covariates may not follow a strictly linear relationship with the log hazard
ratio.

We use the \texttt{cph} function included in the \texttt{rms} package
of the R statistical language to fit a Cox PH model with the 
formulation presented in Listing 2.

Initially, we attempted to fit a single model to all data sampled from the eight sites under study, introducing an additional multilevel categorical parameter to
identify the effect of each site. However, model results and diagnostics showed
a very poor fit. This suggested the alternative of fitting an individual Cox PH
model for each site, as not all independent parameters may exert the same influence
for all sites.

\begin{lstlisting}[breaklines=true,
caption={Model formulation to fit a Cox PH model for each site using the 
\texttt{cph} function in the \texttt{rms} package in R.},
basicstyle=\ttfamily\scriptsize,
frame=lines,
showtabs=false,
showspaces=false,
showstringspaces=false,
captionpos=b]
> cph(formula = Surv(tanswer, solved) ~ rcs(zscore, 3) +
  rcs(log(bodylength), 3) + rcs(log(titlelength), 3) + 
  hasexample + tagscount + rcs(log(sumpeople), 3),
  data = data_stackow, method = "efron", x=T, y=T, surv=T)
\end{lstlisting}

This approach obtained much better diagnostics than a single model for all sites.
Of particular importance is testing the validity of the proportional hazards
assumption, which can be checked through plots of the scaled Schoenfeld
residuals for each parameter against time~\cite{mills2011}. If this assumption
holds, the plots must not show any pattern for residuals over time. In addition,
an individual test is computed to evaluate evidence of a non-null correlation
coefficient of these residuals over time. Listing 3 presents the results of this
test for the \textit{apple} community, obtained with the \texttt{cox.zph} function
in the \texttt{survival} library for R. We confirm that there is no strong evidence
to discard the null hypothesis of a correlation coefficient $rho=0$ for any
parameter in the model with time. Similar successful diagnostics were obtained for all 
remaining Cox PH models in this analysis.

\begin{lstlisting}[breaklines=true,
caption={Model formulation to fit a Cox PH model for each site using the 
\texttt{cph} function in the \texttt{rms} package in R.},
basicstyle=\ttfamily\scriptsize,
frame=lines,
showtabs=false,
showspaces=false,
showstringspaces=false,
captionpos=b]
                   rho   chisq        p
zscore        -0.00777   0.170    0.680
bodylength    -0.01705   0.999    0.318
titlelength    0.00687   0.162    0.687
hasexample    -0.02363   1.944    0.163
tagscount     -0.01767   1.069    0.301
sumpeople      0.00639   0.138    0.711
GLOBAL              NA   6.779    0.342
\end{lstlisting}

Table~\ref{tab:hazard-ratio-estim} summarizes multilevel estimations (combining
contributions from linear and non-linear components) of the hazard ratio for each 
covariate considered in our model. For continuous parameters, the hazard ratio is 
calculated between the upper and lower values of the inter-quartile range, controlling
for all other covariates. For the categorical parameter \texttt{hasexample}, it is
computed between the two possible categories, again controlling for all
other covariates.

\begin{table*}
  \centering
  \begin{tabular}{|c|c|c|c|c|c|c|c|c|}
    \hline
    \tabhead{ } &
    \multicolumn{1}{|p{0.15\columnwidth}|}{\centering\tabhead{Apple}} &
    \multicolumn{1}{|p{0.2\columnwidth}|}{\centering\tabhead{AskUbuntu}} &
    \multicolumn{1}{|p{0.15\columnwidth}|}{\centering\tabhead{Math}} &
    \multicolumn{1}{|p{0.2\columnwidth}|}{\centering\tabhead{ServerFault}} &
    \multicolumn{1}{|p{0.2\columnwidth}|}{\centering\tabhead{SharePoint}} &
    \multicolumn{1}{|p{0.2\columnwidth}|}{\centering\tabhead{StackOverflow}} &
    \multicolumn{1}{|p{0.17\columnwidth}|}{\centering\tabhead{SuperUser}} &
    \multicolumn{1}{|p{0.17\columnwidth}|}{\centering\tabhead{Wordpress}}\\
    \hline
    zscore & \cellcolor[HTML]{F6CECE}0.96 & \cellcolor[HTML]{F6CECE}0.92 & 0.97 & \cellcolor[HTML]{F6CECE}0.95 & 0.98 & 0.98 & \cellcolor[HTML]{F6CECE}0.97 & \cellcolor[HTML]{F6CECE}0.96\\
    \hline
    bodylength & \cellcolor[HTML]{F78181}0.79 & \cellcolor[HTML]{F78181}0.76 & \cellcolor[HTML]{F78181}0.77 & \cellcolor[HTML]{F78181}0.80 & \cellcolor[HTML]{F78181}0.81 & \cellcolor[HTML]{F78181}0.74 & \cellcolor[HTML]{F78181}0.77 & \cellcolor[HTML]{F78181}0.80\\
    \hline
    titlelength & \cellcolor[HTML]{F6CECE}0.94 & 1.04 & 1.04 & 0.96 & \cellcolor[HTML]{F6CECE}0.93 &  \cellcolor[HTML]{F6CECE}0.94 & 0.98 & \cellcolor[HTML]{F6CECE}0.94\\
    \hline
    tagscount & 0.95 & 1.01 & \cellcolor[HTML]{F78181}0.64 & \cellcolor[HTML]{F6CECE}0.89 & \cellcolor[HTML]{F6CECE}0.94 &  \cellcolor[HTML]{F6CECE}0.83 & 0.94 & \cellcolor[HTML]{F6CECE}0.94\\
    \hline
    sumpeople & 1.09 & 0.98 & \cellcolor[HTML]{97CDAB}1.44 & \cellcolor[HTML]{CDFBDF}1.28 & 1.09 & \cellcolor[HTML]{97CDAB}1.46 & 1.08 & 1.07\\
    \hline
    hasexample & 0.98 & \cellcolor[HTML]{CDFBDF}1.22 & 0.94 & 0.96 & \cellcolor[HTML]{CDFBDF}1.19 & \cellcolor[HTML]{F78181}0.70 & 1.06 & \cellcolor[HTML]{F6CECE}0.89\\
    \hline
  \end{tabular}
  \caption{Estimators of the hazard ratio corresponding to each covariate considered
  in the Cox PH model. Green colours indicate positive effects (increasing hazard ratio),
  whereas red colours mark negative effects (decreasing hazard ratio).}
  \label{tab:hazard-ratio-estim}
\end{table*}

Interpreting the statistical significance of each covariate on the hazard ratio can be misleading if we base our considerations solely on traditional significance tests for individual model coefficients and the associated p-values. In situations like this, where
we have both linear and non-linear components (from the restricted cubic splines transformation), it is non-trivial to interpret p-values calculated for each individual component. Furthermore, p-values can be misleading with large samples, as the test statistic is inversely proportional to the $\sqrt{n}$, being $n$ the size of the sample. Since we used large samples ($n \sim 5,000$), it becomes harder to know if what we are observing are true effects or just noise.

\begin{figure*}
\vspace{-1cm}
\centering
\subfloat[Apple]{
  \includegraphics[width=0.98\columnwidth,height=5cm]{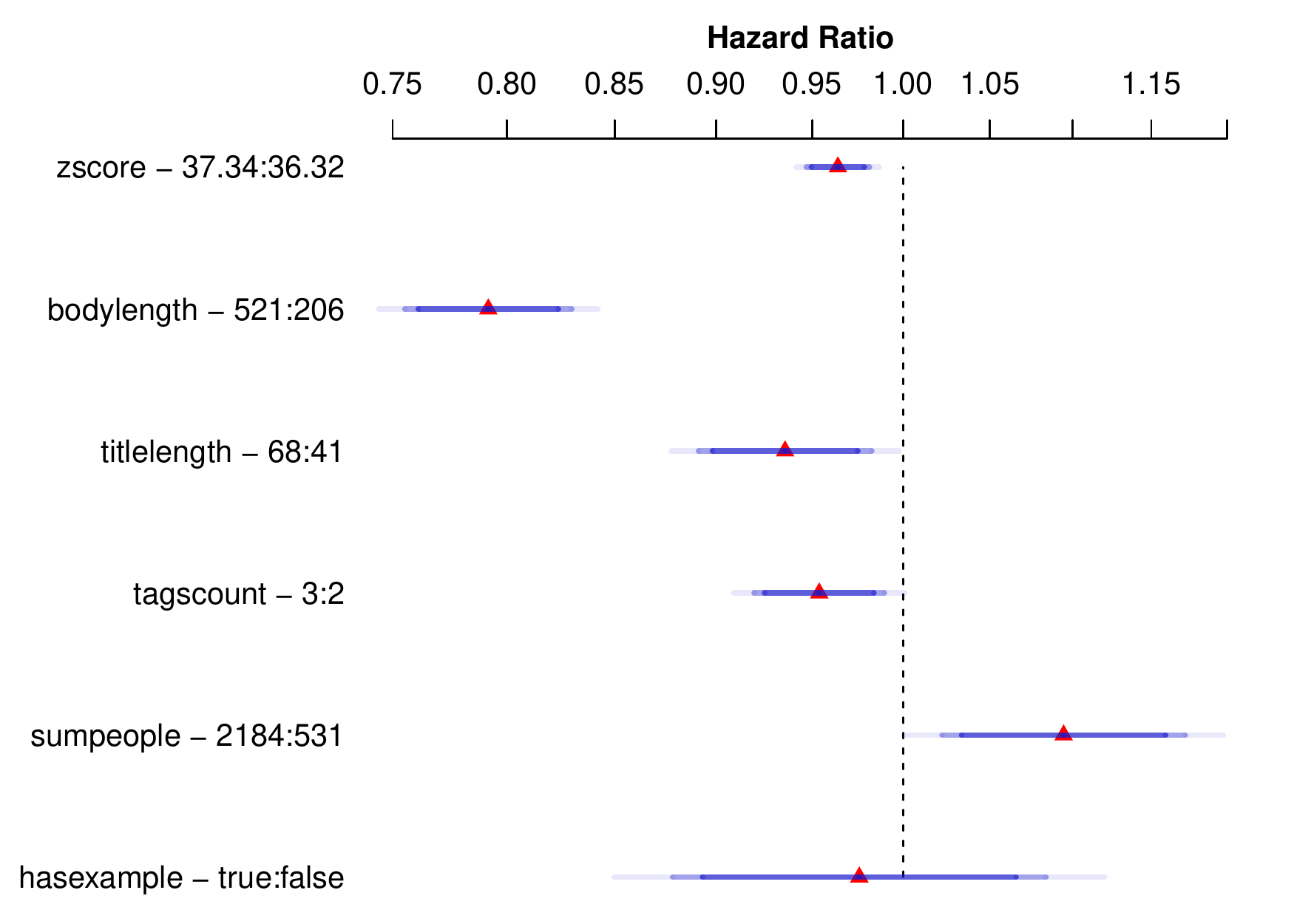}
}
\subfloat[AskUbuntu]{
  \includegraphics[width=0.98\columnwidth,height=5cm]{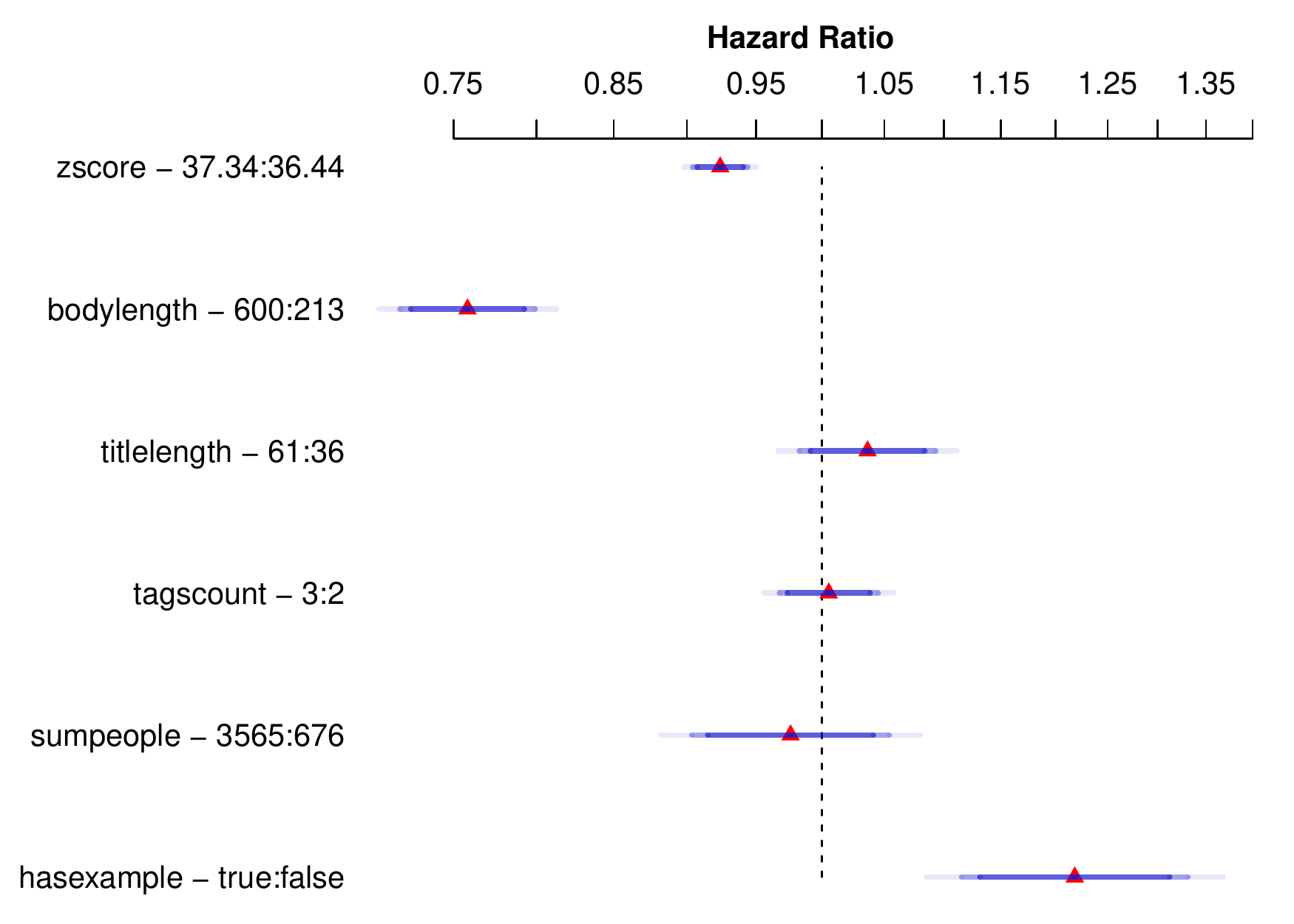}
}
\hspace{0mm}
\subfloat[Math]{
  \includegraphics[width=0.98\columnwidth,height=5cm]{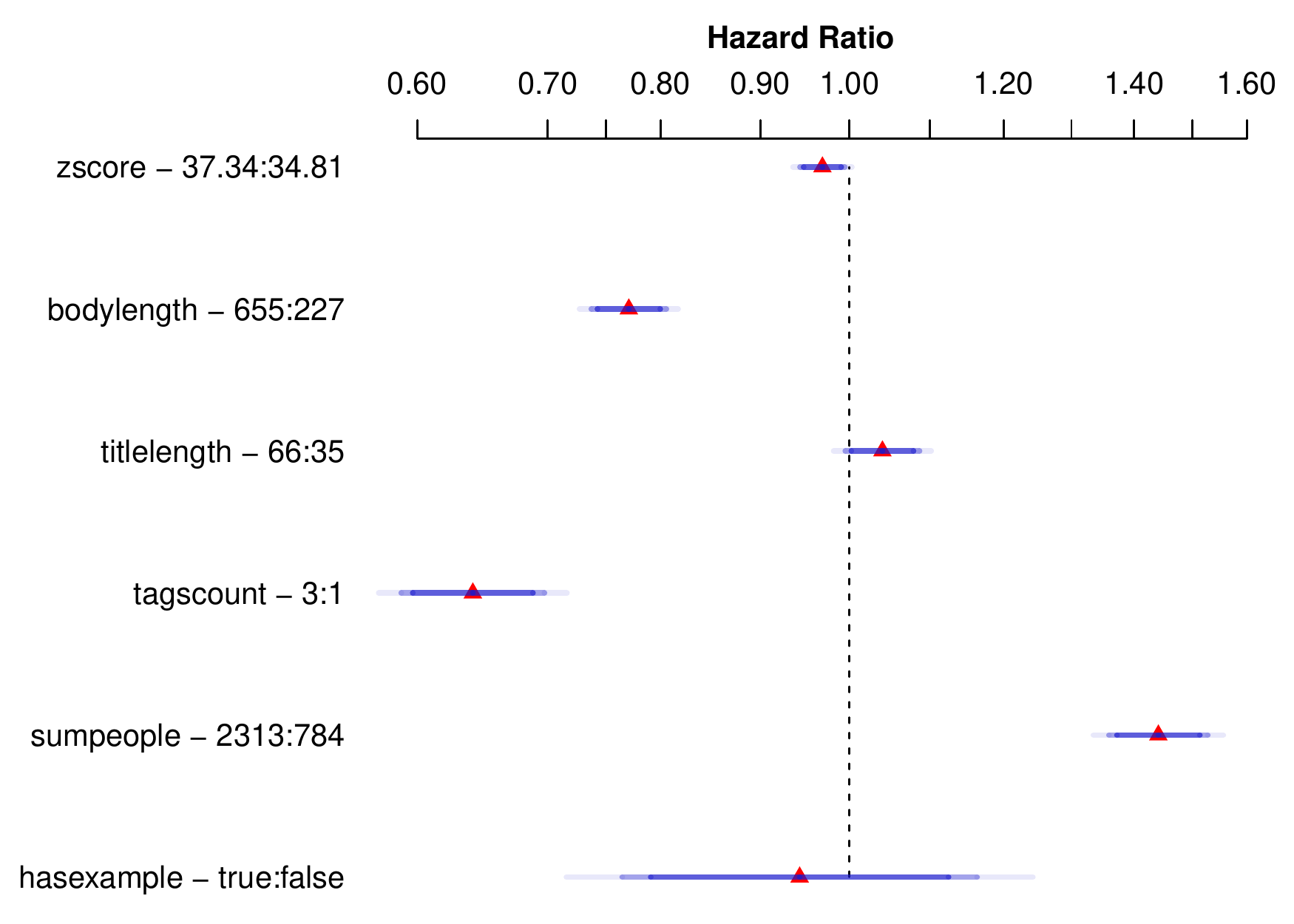}
}
\subfloat[ServerFault]{
  \includegraphics[width=0.98\columnwidth,height=5cm]{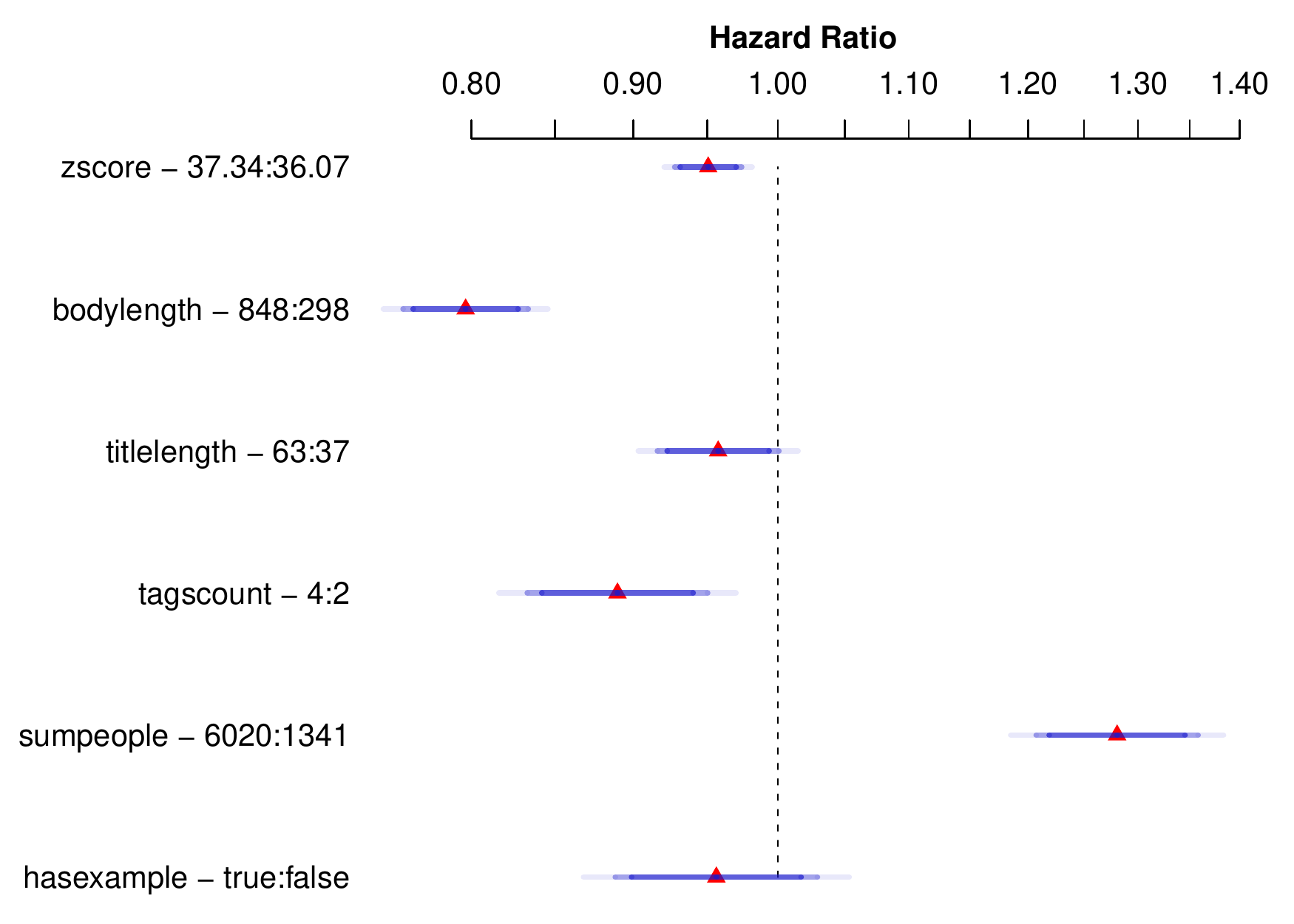}
}
\hspace{0mm}
\subfloat[SharePoint]{
  \includegraphics[width=0.98\columnwidth,height=5cm]{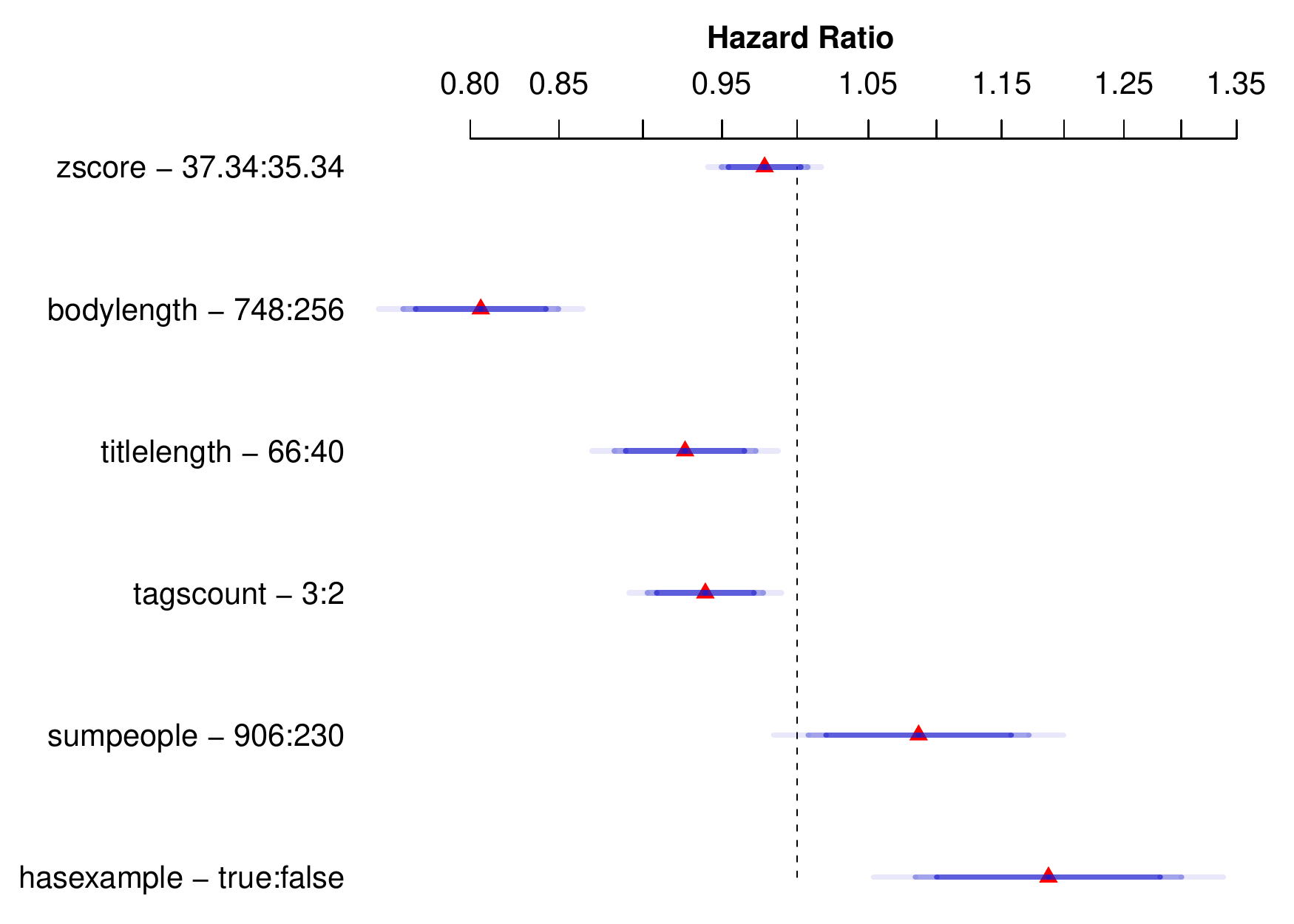}
}
\subfloat[StackOverflow]{
  \includegraphics[width=0.98\columnwidth,height=5cm]{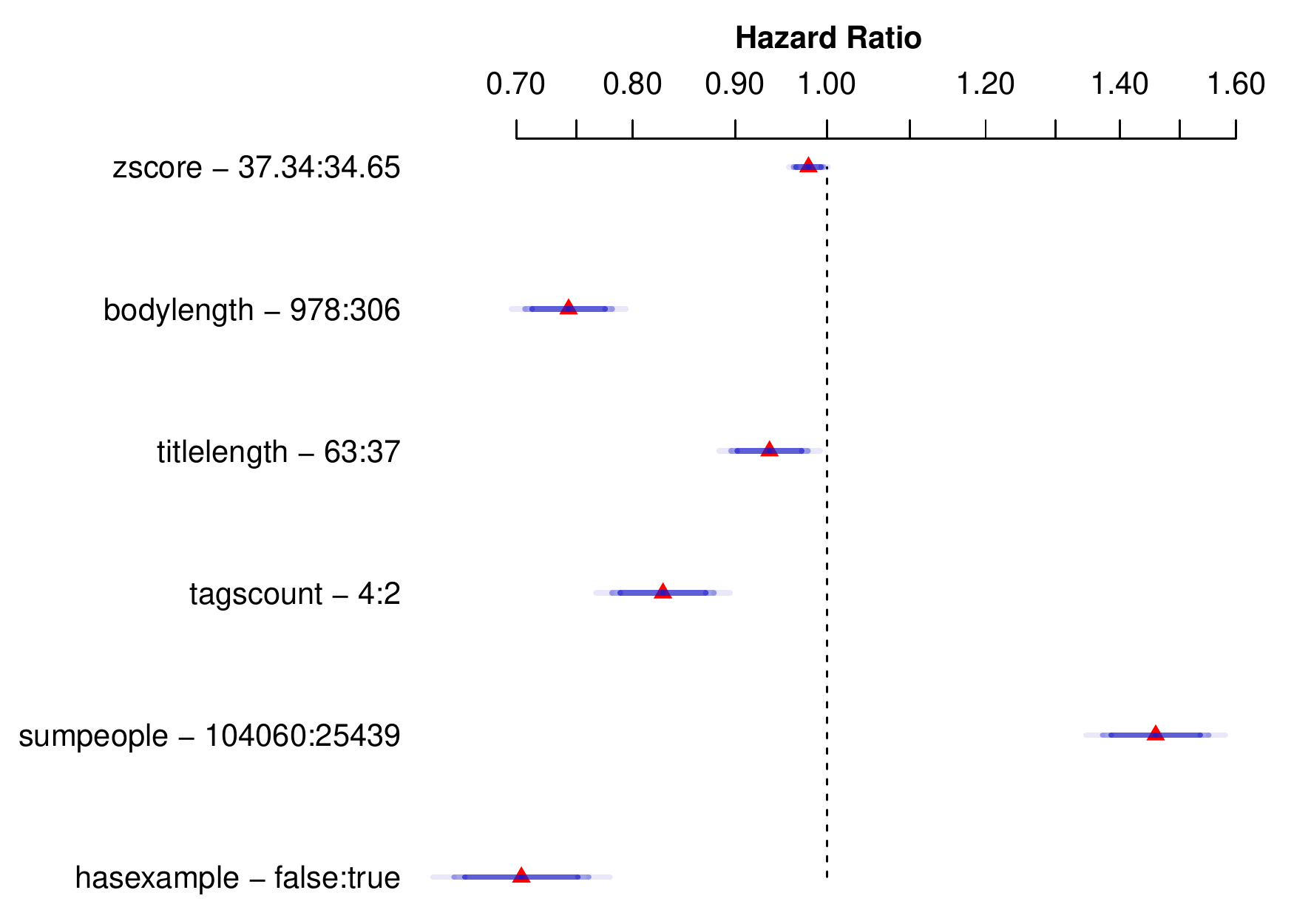}
}
\hspace{0mm}
\subfloat[Superuser]{
  \includegraphics[width=0.98\columnwidth,height=5cm]{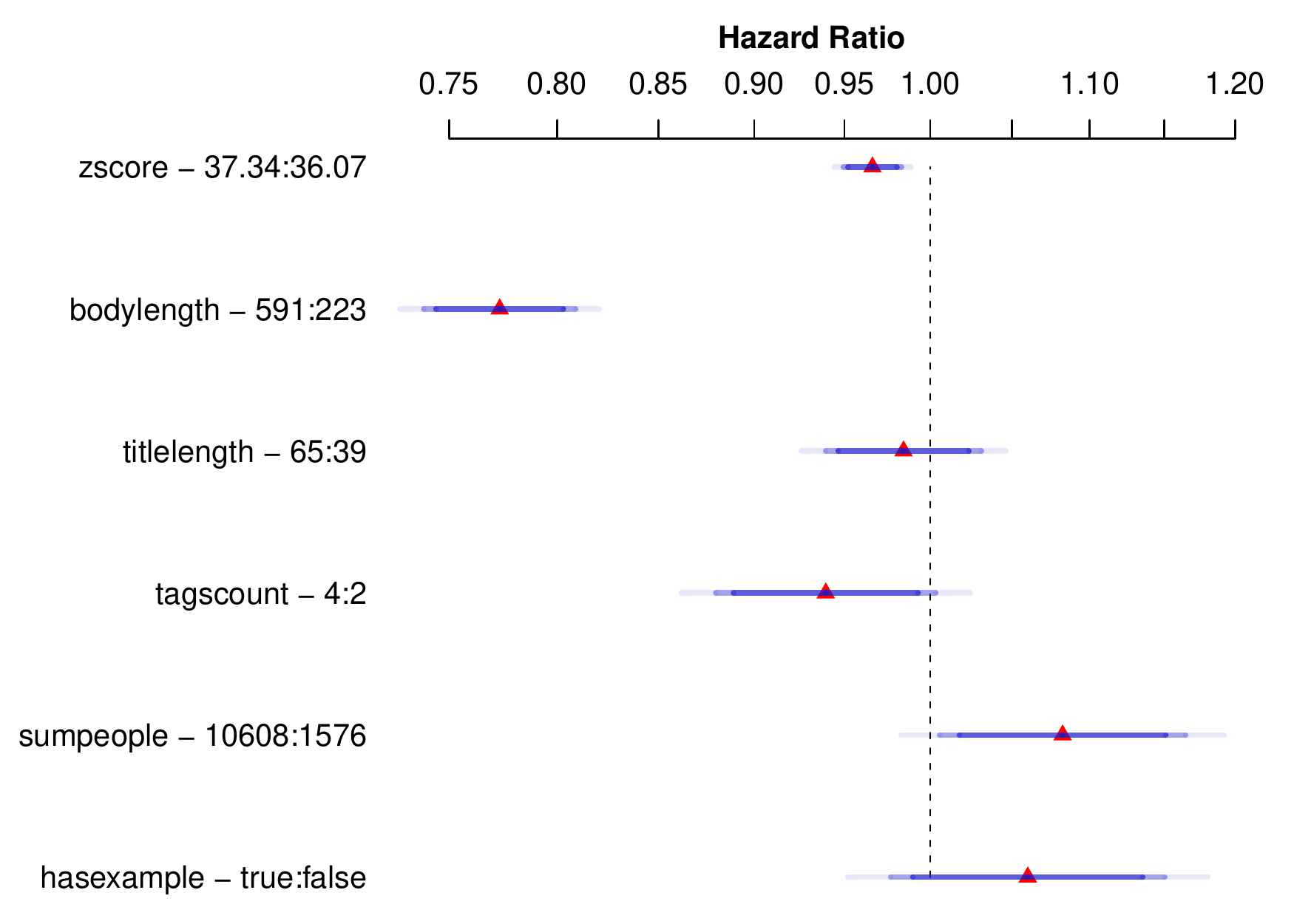}
}
\subfloat[Wordpress]{
  \includegraphics[width=0.98\columnwidth,height=5cm]{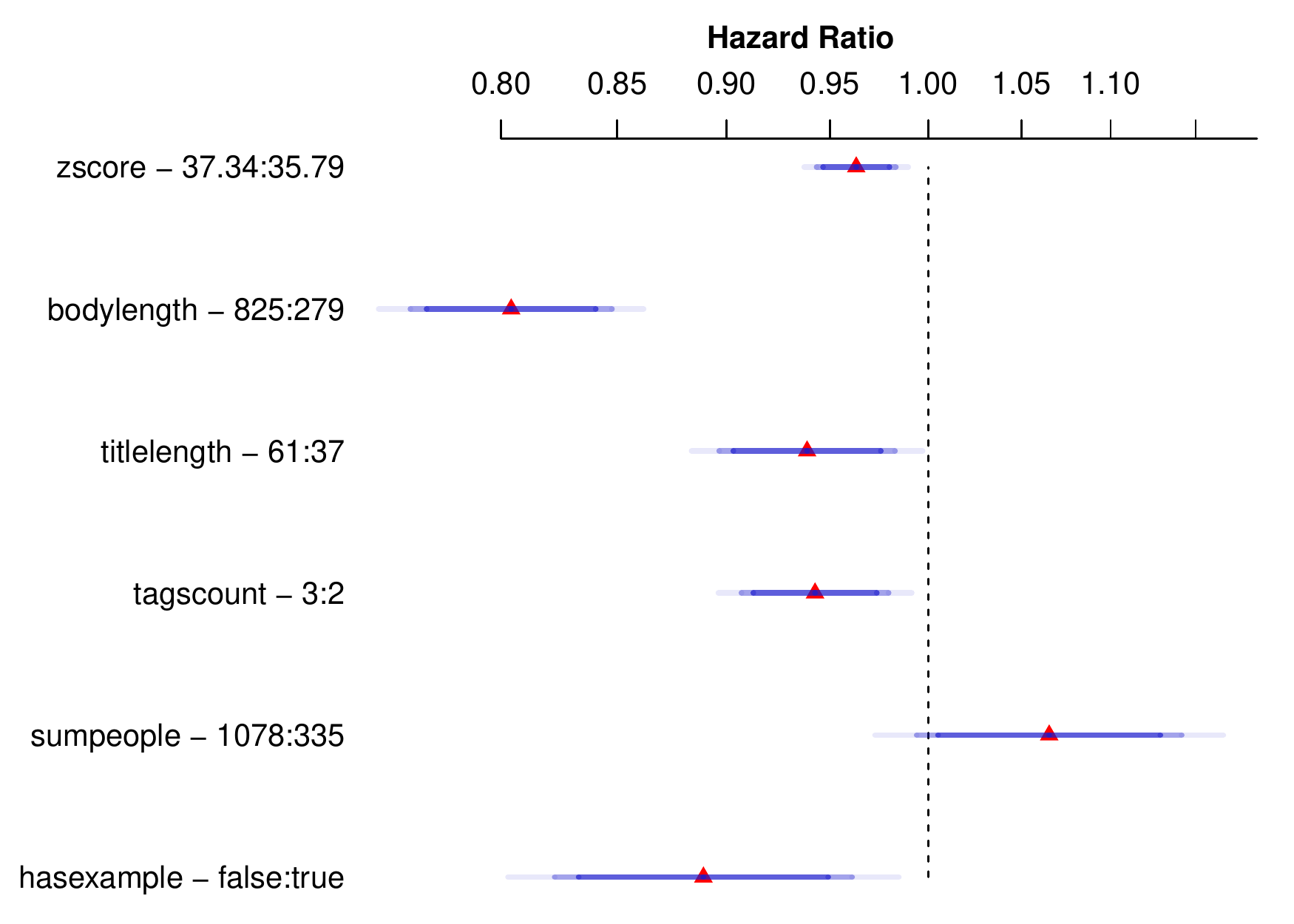}
}
\caption{Estimated hazard ratios and multilevel confidence intervals at 90\% (dark blue),
95\% (blue) and 99\% (light blue) for model covariates in each site. 
Hazard ratios are calculated between the lower and upper 
limits of the interquartile range (continuous variables) or comparing between 
levels (categorical variables), as indicated next to the label of each covariate.}
\label{fig:multilevel-cis}
\end{figure*}

To avoid these issues, we preferred to report multilevel confidence intervals (adjusted to combine linear and non-linear components in our model). By doing so, we follow modern approaches in reporting and assessment of statistical models~\cite{cumming2013}.
Figure~\ref{fig:multilevel-cis} shows such a plot, created with the \texttt{rms}
package in R. For each point estimator of the hazard ratio (marked with a red
triangle) confidence intervals are plotted at levels 0.9, 0.95 and 0.99. The
vertical dashed line marks the unity value for the hazard ratio (no effect).
In consequence, if the confidence intervals include the dashed line we do not
have strong empirical evidence of a significant effect for the corresponding
covariate. Otherwise, there is strong evidence supporting the claim for such
an effect of that covariate on the hazard ratio.

\begin{figure*}[t]
\centering
\includegraphics[width=1.75\columnwidth]{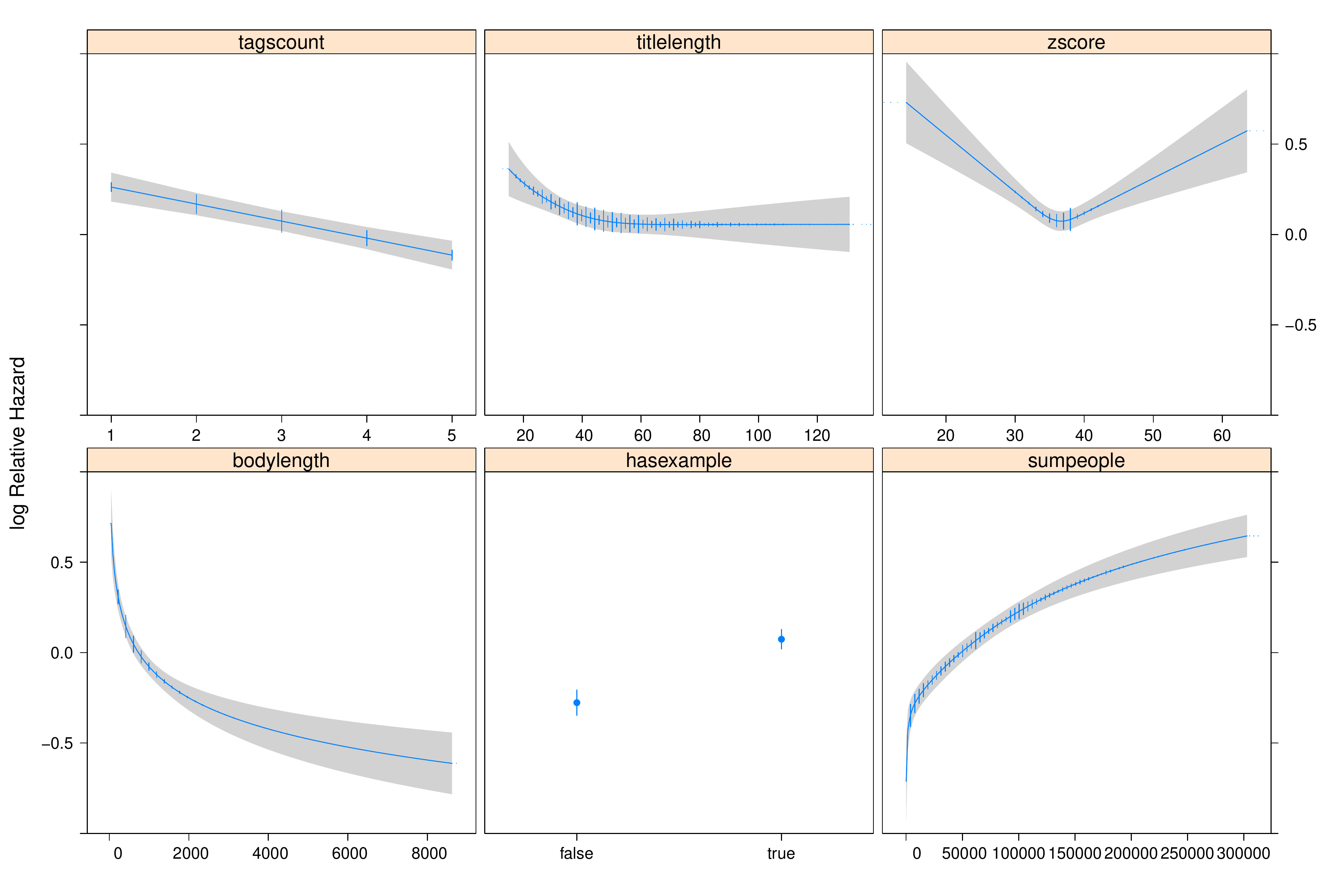}
\caption{Plots of effects of covariates included in the model on the log hazard ratio for the \texttt{stackoverflow} site.}
\label{fig:effects}
\end{figure*}

Finally, in addition to the point estimators for effects on the hazard ratio presented
above, it is also possible to create effect plots showing the influence of each
covariate along a range of values, controlling for all other covariates. This is also
a very useful feature shipped in the \texttt{rms} package in R. Figure~\ref{fig:effects}
show these plots for each covariate in our model in the case of the \textit{stackoverflow}
site. Besides, these graphs are already adjusted to plot the effects on the original
scale of each covariate in the model, as the package can 'remember' the transformations
that were applied to each parameter. 95\% confidence intervals are also presented in
shaded grey along each effect curve.

This type of graph allows to visually interpret the shape of effects on the
outcome variable. For example we can see that the strong effect reported for the
\texttt{bodylength} parameter presents a smoothed declining shape. For very short
values of \texttt{bodylength} (short questions) the log hazard ratio increases and thus
the chances of a question to be resolved. However, questions with \texttt{bodylength}
values greater than 1 KB have a negative log hazard ratio, indicating a lower chance
of being resolved. Likewise, we can confirm that questions including an example increase
their chances to be resolved, but those without an example decrease their likelihood
of being resolved. Finally, we need more than 50,000 users subscribed to the tags
used to label the question to increase it chance of being resolved, and this effect
grows rapidly as the size of the audience increases. This provides empirical evidence
for a \textit{wisdom of the crowd} effect~\cite{surowiecki2005} in \textit{StackOverflow}:
the larger the audience, the higher the chance of resolving our question.

\section{Implications for tool design and management}

Since a survival model built along the lines discussed above may indicate the likelihood 
of a given question to receive an answer and provide a reasonable estimation of the 
time it will take, these aspects may be used to improve QA platforms in different 
ways: by either providing an automatic feedback to the questioner or, as in the hybrid design proposed in our previous work \cite{piccardi2014}, to merge crowdsourcing with an expert-based model to improve the efficacy of the service.

A first important lesson that emerges from the results is that a 
one-model-fits-all approach is less effective than a site-specific approach in building predictive models. This lesson, which holds in the context of the features that we have selected (see also further 
discussion in the conclusion section), suggests a few implications for the design of QA platforms.

In the first place, to best predict question resolution site by site, software platforms that support multiple QA sites (and involving humans via mixed-initiative user interfaces) should surface unique subsets of predictive features within the automatic predictor (i.e., to inform the human) and the user interface (i.e., to guide the human as s/he acts on the question) (e.g., see \cite{piccardi2014}).
Moreover, even when two QA sites share the same set of predictive feature, the underlying QA software platform should accommodate unique prioritization of the subsets of features within the automatic predictor and the user interface: i.e., two different QA sites running on the same platform with the same set of predicting features may surface them with different priorities; the same site that changes significantly over time may surface the features with different priorities along its lifecycle.

The second implication for design may even bring us to a more ambitious vision to move from systems that are first designed, then used, and eventually redesigned, to systems that allow for QA site \textit{use and adaptive redesign to run in parallel} as the underlying QA platform detects from the site usage history what are its \textit{current} subset of predictors (significant estimators) and model coefficients (size of effects per estimator).  Indeed, a limitation seldom discussed of accurate modelling of services is the case when they are used to improve 
the service, whose technology infrastructure might become outdated by the time is design, tested, and deployed.

\section{Conclusions and future work}

In this paper, we have used survival analysis to model the time to answer questions posed on a wide set of different QA sites taken from the same QA platform.  Since the test for PH assumption holds in our analysis, survival models are valid and can be used to predict 
the speed to answer questions in a QA site that is comparable to those we analyzed.

We explicitly decided not include content-based analysis in order to improve portability of the models across different QA domains (e.g. programming vs. Math). Of course, this is at the same time a limitation and an opportunity for future work since content-based features are likely to improve accuracy of models~\cite{nasehi2012}.

Furthermore, we have driven our feature set from a literature survey and there is no direct evidence that a different set of features cannot be more useful. Although this aspect may be part of our future work, we contend that our features are good enough for a baseline model and that they are easy to compute. Further work is definitely needed to provide 
generalized figures of merits.

Overall, we believe that the present study shed some new light on the field of crowd-based QA sites and it provides a new approach through survival analysis to model one of the crucial aspects in this field, namely to determine the likelihood of a question to receive an answer and an estimation of the time required to receive the accepted answer that solves that question.

%
%
%
%
%
\balance

\bibliographystyle{acm-sigchi}
\bibliography{survival-qa}
\end{document}